\documentclass[aps,twocolumn,prl,showpacs]{revtex4}

\usepackage{graphicx}
\usepackage{dcolumn}
\usepackage{bm}

\newcommand{\dograph}[2]{
	\begin{figure}
	\includegraphics[width=3.375in]{#1}
	\caption{\label{fig:#1}#2}
	\end{figure}
}

\begin{document}

\title{Anomalous modulation of a zero bias peak in a\\hybrid nanowire-superconductor device}

\author{A.D.K.~Finck$^1$, D.J.~Van Harlingen$^1$, P.K.~Mohseni$^2$, K.~Jung$^2$, X.~Li$^2$}
\affiliation{$^1$Department of Physics and Materials Research Laboratory, University of Illinois at Urbana-Champaign, Urbana, Illinois 61801
\\
$^2$Department of Electrical and Computer Engineering, Micro and Nanotechnology Laboratory, University of Illinois at Urbana-Champaign, Urbana, Illinois 61801}

\date{\today} 

\begin{abstract}
We report on transport measurements of an InAs nanowire coupled to niobium nitride leads at high magnetic fields.  We observe a zero-bias anomaly (ZBA) in the differential conductance of the nanowire for certain ranges of magnetic field and chemical potential.  The ZBA can oscillate in width with either magnetic field or chemical potential; it can even split and re-form.  We discuss how our results relate to recent predictions of hybridizing Majorana fermions in semiconducting nanowires, while considering more mundane explanations.
\end{abstract}


\maketitle

Majorana fermions (MFs) are neutral particles that are their own antiparticles.  Although they were first proposed to describe fundamental particles\cite{Majorana1937}, recent years have seen intense interest in realizing solid-state systems with quasiparticles that behave like MFs \cite{Beenakker2011, Alicea2012}.  There are several candidates, including certain quantum Hall states \cite{PhysRevB.61.10267} and topological insulators coupled with superconductors \cite{PhysRevLett.100.096407}.  Solid-state MFs can be used to create a topological quantum computer, in which the non-Abelian exchange statistics of the MFs are used to process quantum information non-locally, evading error-inducing local perturbations \cite{PhysRevLett.94.166802,RevModPhys.80.1083}.

A promising candidate is a one-dimensional spinless $p$-wave superconductor \cite{kitaev2001}.  One can engineer this system in a semiconductor nanowire with strong spin-orbit coupling \cite{PhysRevLett.105.077001, PhysRevLett.105.177002}, which separates the two electron helicities in energy.  Applying a Zeeman splitting perpendicular to the spin-orbit coupling can create an energy range where only one helicity is present, effectively generating a spinless system.  If superconductivity is induced by an $s$-wave superconductor, Pauli exclusion will require the nanowire to acquire $p$-wave pairing symmetry.  This proposal is attractive because supercurrents have already been observed in InAs nanowires \cite{Science.309.272, JApplPhys.112.034316}.  A nanowire with a single occupied subband goes from the spinful to the spinless regime when $E_Z^2 > \mu^2 + \Delta^2$, where $E_Z=\frac{1}{2} g \mu_B B$ is the Zeeman energy, $\mu$ is the chemical potential defined relative to the bottom of the subband, and $\Delta$ is the induced superconducting pairing.  When passing between these two regimes, the nanowire undergoes a topological phase transition in which the single-particle gap collapses and changes sign.  If a nanowire has a spinless (i.e. topological) segment in between two spinful (i.e. trivial) segments, then the nanowire will harbor a single pair of MFs that exist as zero energy modes pinned to the boundaries separating the topologically distinct regions.  Although disorder \cite{PhysRevB.84.144526, PhysRevLett.109.146403, PhysRevB.84.195436}, Coulomb interactions \cite{PhysRevB.84.180502}, and multiple subbands \cite{PhysRevLett.106.127001, PhysRevB.84.144522, PhysRevB.84.214528} might quantitatively change the conditions for MFs, the qualitative picture should remain: for certain ranges of parameters the nanowire will be in the topological regime and contain a pair of MFs.

A key probe for MFs is tunneling spectroscopy \cite{PhysRevB.63.144531, PhysRevLett.103.237001, PhysRevB.82.180516, PhysRevB.82.214509, PhysRevLett.109.227006}.  The MF would manifest as a conductance peak at zero voltage.  The MFs can only interact with other MFs, so the peak would stay at zero so long as the MFs are spatially separated from each other.  Indeed, numerous groups \cite{Mourik25052012, Deng2012, NPhysics2479, arXiv.1211.3954} have reported zero bias anomalies (ZBAs) in devices inspired by the theoretical proposals.  However, a ZBA might also occur under similar conditions due to a Kondo resonance that manifests when the magnetic field has suppressed the superconducting gap enough to permit the screening of a localized spin \cite{PhysRevLett.109.186802}.  Thus, it is necessary to seek more definitive signatures of MFs.

One possibility is to look for signs that the MFs are hybridizing with each other \cite{PhysRevLett.103.107001, PhysRevB.82.094504, pikulin2011,PhysRevB.86.180503, arXiv.1207.5907, PhysRevB.86.220506}.  Because the wave functions of MFs decay exponentially within the interior of the topological nanowire, MFs at the ends of a finite nanowire will overlap with each other and hybridize.  The hybridization magnitude can be tuned by the Zeeman energy or chemical potential, which control the decay length of the MF wave function and the period of its oscillatory component.  The ZBA would then split and re-form in a periodic fashion, in contrast with the linear splitting expected for the Kondo effect.

In this Letter, we report on the behavior of ZBAs in an InAs nanowire coupled to superconducting leads.  We focus on the regime of large magnetic fields to suppress extraneous effects, including Josephson supercurrents, Kondo resonances \cite{Nature.391.156, PhysRevB.84.245316}, and reflectionless tunneling \cite{PhysRevLett.69.510}.  We find that the ZBAs are robust against changes in Zeeman energy and chemical potential.  Under certain conditions, the width of the ZBA oscillates with either parameter.  The ZBA can even split and re-form.  We argue that this is consistent both with MFs as well as a Kondo effect periodically generated by resonant levels.

Metal-organic chemical vapor deposition growth of InAs nanowires with 100 nm diameters was performed using the Au-assisted vapor-liquid-solid approach.  TEM analysis reveals a nearly purely hexagonal (wurtzite) crystal structure, with only a few stacking faults near the nanowire tips.  The nanowires are then deposited on a Si substrate with a 300 nm thick SiO$_2$ dielectric, permitting back gating.  Superconducting leads are defined by conventional e-beam lithography.  The leads overlapping the nanowire are each 1 $\mu$m wide and separated by 150 nm.  We sputter 55 nm of niobium nitride via a DC sputter gun and a Nb target in an Ar environment with a partial pressure of N$_2$.  Immediately prior to sputtering, the contact regions are briefly exposed to an Ar ion mill to remove the native oxide and permit transparent contacts \cite{Nanotechnology.22.445701}.  We note that ion milling can raise the carrier density within the contact region with respect to the unetched InAs nanowire \cite{APL.67.3569}.  The NbN thin film has a critical temperature of 12 K and an upper critical field of 9 T at 10 K.  The sample is lowered into the mixing chamber of a top-loading dilution refrigerator.  Immersion in the dilute phase of the mixture provides an excellent thermal sink as evidenced by the continuing evolution of the transport measurements below 50 mK.  The I-V characteristics of the superconductor/nanowire/superconductor junction are measured via standard lockin techniques, employing a 10 $\mu$V AC excitation at 73 Hz.  Unless otherwise stated, all reported data were taken at a mixing chamber temperature of 10 mK.  To induce Zeeman splitting, we apply a magnetic field perpendicular to the Si substrate.

\dograph{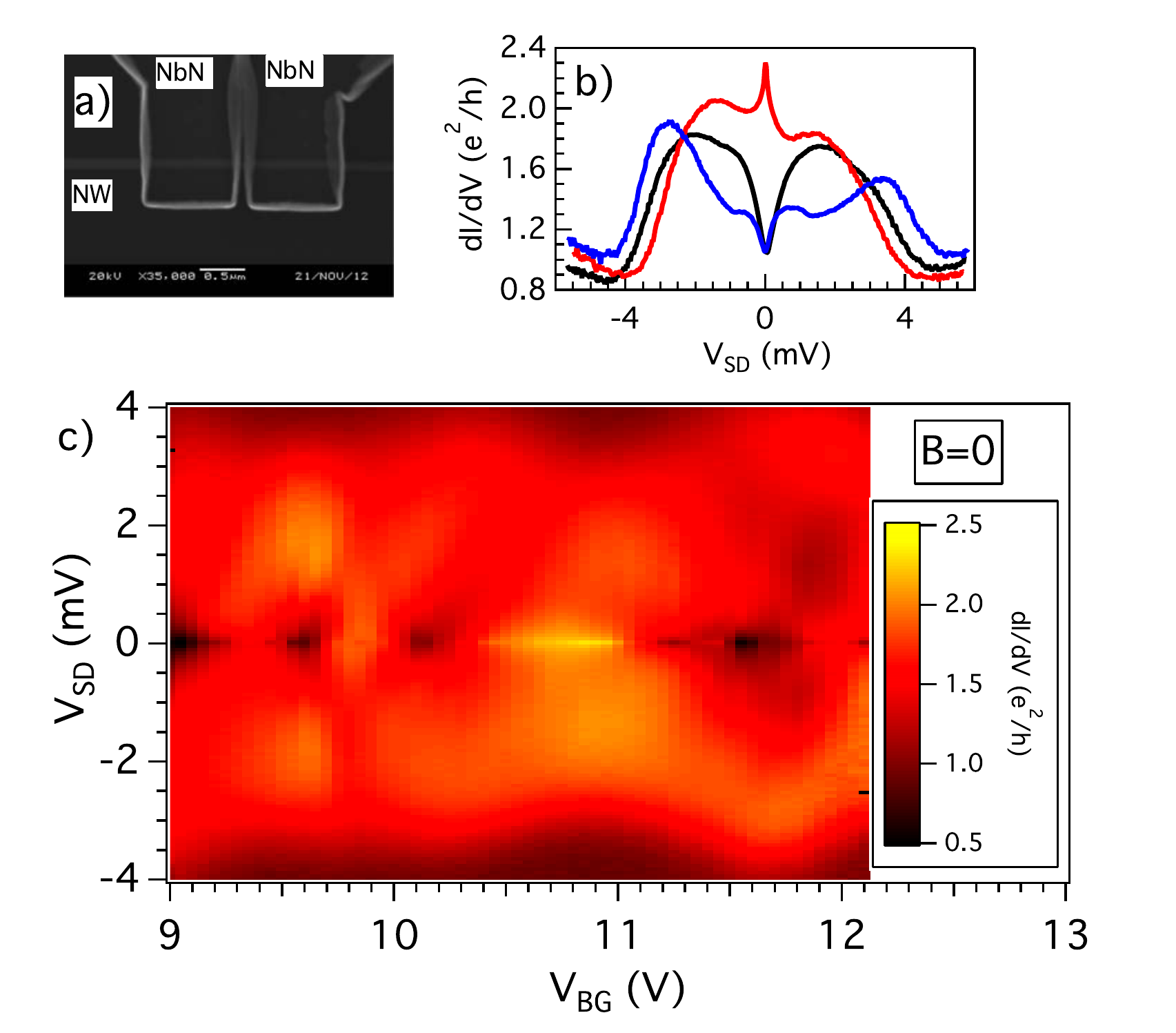}{(a) SEM image of NbN leads on InAs nanowire (NW).  The white feature at the edges of the NbN leads is PMMA residue from the ion milling.  (b) Transport at $B=0$ for $V_{BG} = 10.1$ V (black), 10.9 V (red), and 11.75 V (blue).  (c) Stability diagram at $B=0$ for region of interest.}

To demonstrate that a superconducting proximity effect is induced in the nanowire, we first consider transport through the nanowire at zero magnetic field.  We focus on the density range where we observe a rough plateau in the high bias conductance of approximately $e^2/h$, suggesting that there is one occupied subband with a transparency of 0.5.  Representative conductance curves are shown in Figure 1b.  We observe an enhancement of the differential conductance for source-drain voltage $|V_{SD}| < 4$ mV by a factor of approximately 2 beyond the high bias conductance.  Our NbN films have a gap of $\Delta_0 \approx 2$ meV, suggesting that we are observing Andreev reflection at the transparent nanowire-superconductor interface for $|V_{SD}| < 2 \Delta_0$ \cite{PhysRevB.25.4515}, with an additional voltage drop across the bare portion of the nanowire.

For moderate density, the conductance near $V_{SD}=0$ fluctuates between having either a valley or peak with periodicity of $\Delta V_{BG} \approx 0.6$ V.  This behavior is likely caused by resonant levels within the bare nanowire segment passing through zero energy and allowing transport between the proximitized nanowire segments \cite{PhysRevLett.91.057005}.  This is verified by a checkerboard pattern in the stability diagram that becomes more apparent beyond $V_{BG}=12$ V \cite{Nature.439.953, PhysRevLett.96.207003, NanoLett.10.3439}.  These resonances arise when reflections at the interface between the bare nanowire segment and the NbN-covered segments induce constructive interference, corresponding to the condition $2 k_F L = 2 \pi n$, where $k_F$ is the Fermi wavelength, $L = 150$ nm is the length of the bare segment, and $n$ is an integer.  This is equivalent to the condition $\mu = (\hbar^2 \pi^2 n^2)/(2 m^* L^2)$, where $m^* = 0.023 m_e$ is the electron effective mass.  For small $n$, one expects the resonances to be separated in energy by $\Delta \mu \approx (\hbar^2 \pi^2 n^2)/(2 m^* L^2) \approx 0.7$ meV.  The slopes of stability diagram features suggest the relationship $\Delta \mu = 10^{-3} \Delta V_{BG}$, leading to a predicted back gate periodicity of 0.7 V.  Coupling to the leads can broaden these resonances in terms of energy.

Because the edge of the valleys consistently attains a maximum value of $|V_{SD}| = 600$ $\mu$V despite changes in the structure of the resonant levels, we identify this energy as twice the induced gap, $2\Delta$ \cite{PhysRevB.84.100506}.  Coherence peaks can also be discerned at this voltage for certain ranges of $V_{BG}$, demonstrated by the blue trace in Figure \ref{fig:figure1}b.  When the transmission probability through the bare nanowire segment is low, we observe suppressed conductance for energies below the induced gap of the adjacent segments.  Otherwise, we observe Andreev reflection in this energy range.  However, we note that this energy might instead correspond to the separation of the resonant levels.

As demonstrated in Figures \ref{fig:figure1}b and \ref{fig:figure1}c, at zero magnetic field we see no true supercurrent but we do observe a number of sharp peaks at zero bias.  As $V_{BG}$ changes, the peaks split and become sharp dips, suggesting a complex interplay between superconductivity and the Kondo effect \cite{Morpurgo08101999, PhysRevLett.89.256801, PhysRevLett.99.126602, PhysRevLett.99.136806, PhysRevB.79.161407, PhysRevB.82.054512, PhysRevLett.109.186802}.  At $V_{BG}= 10.9$ V, a zero bias peak is visible, which disappears without signs of splitting beyond a magnetic field of $B=0.4$ T, comparable to the estimated 0.14 T required to have one magnetic flux quantum through the bare nanowire segment. This suggests that the peak results from phase-coherent transport that is broadened by fluctuations.  The peak's critical field is consistent with other Josephson junctions based on semiconducting nanowires with similar dimensions and niobium leads \cite{JApplPhys.112.034316, ApplPhysLett.96.132504}.  Thus, we assert that supercurrent through the nanowire is highly suppressed for $B > 0.4$ T.

\dograph{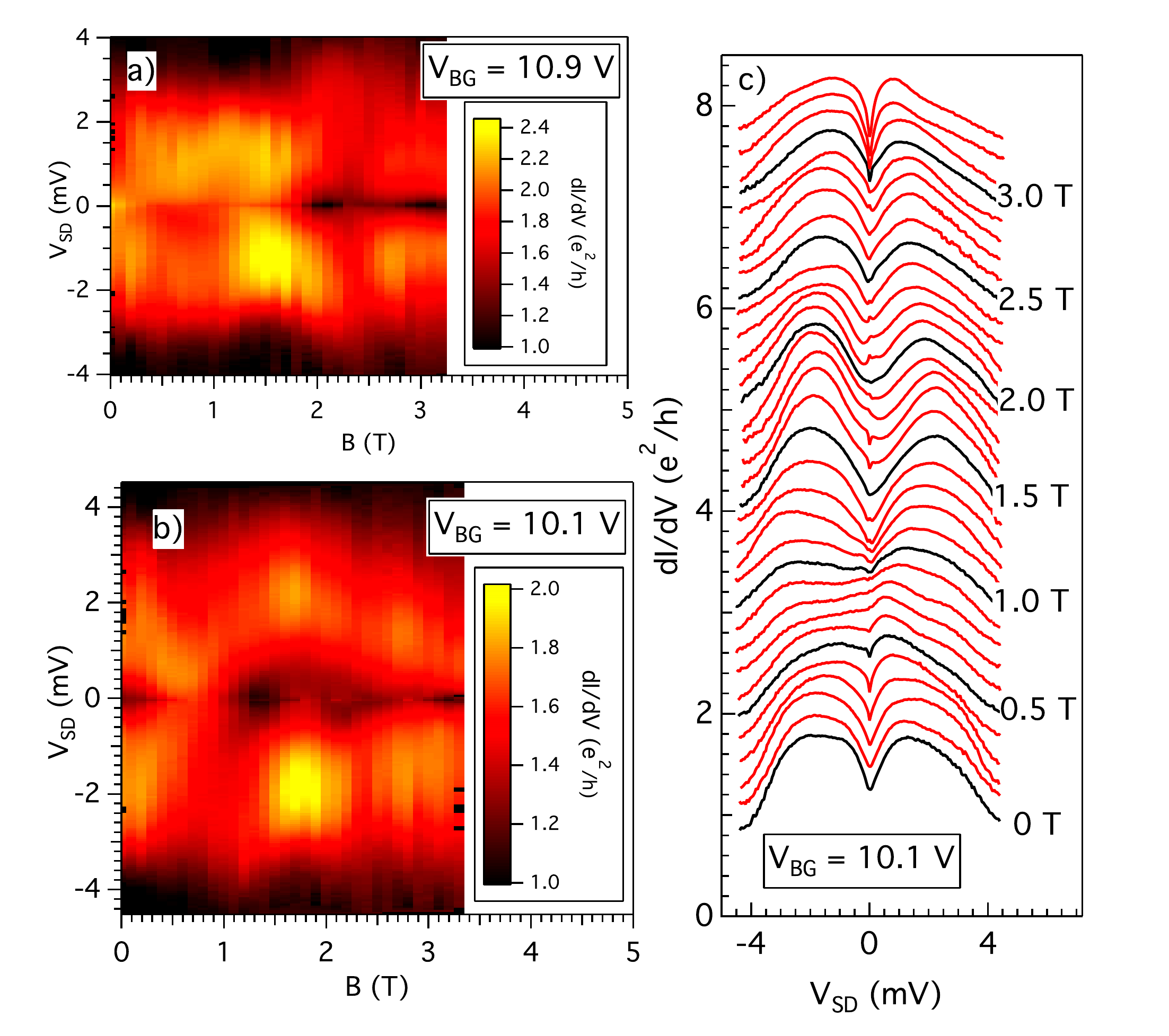}{(a) and (b) Conductance traces vs magnetic field for two different densities.  (c) Individual traces for $V_{BG} = 10.1$ V, each trace differing by 0.1 T.  Adjacent curves are offset for clarity.}

\dograph{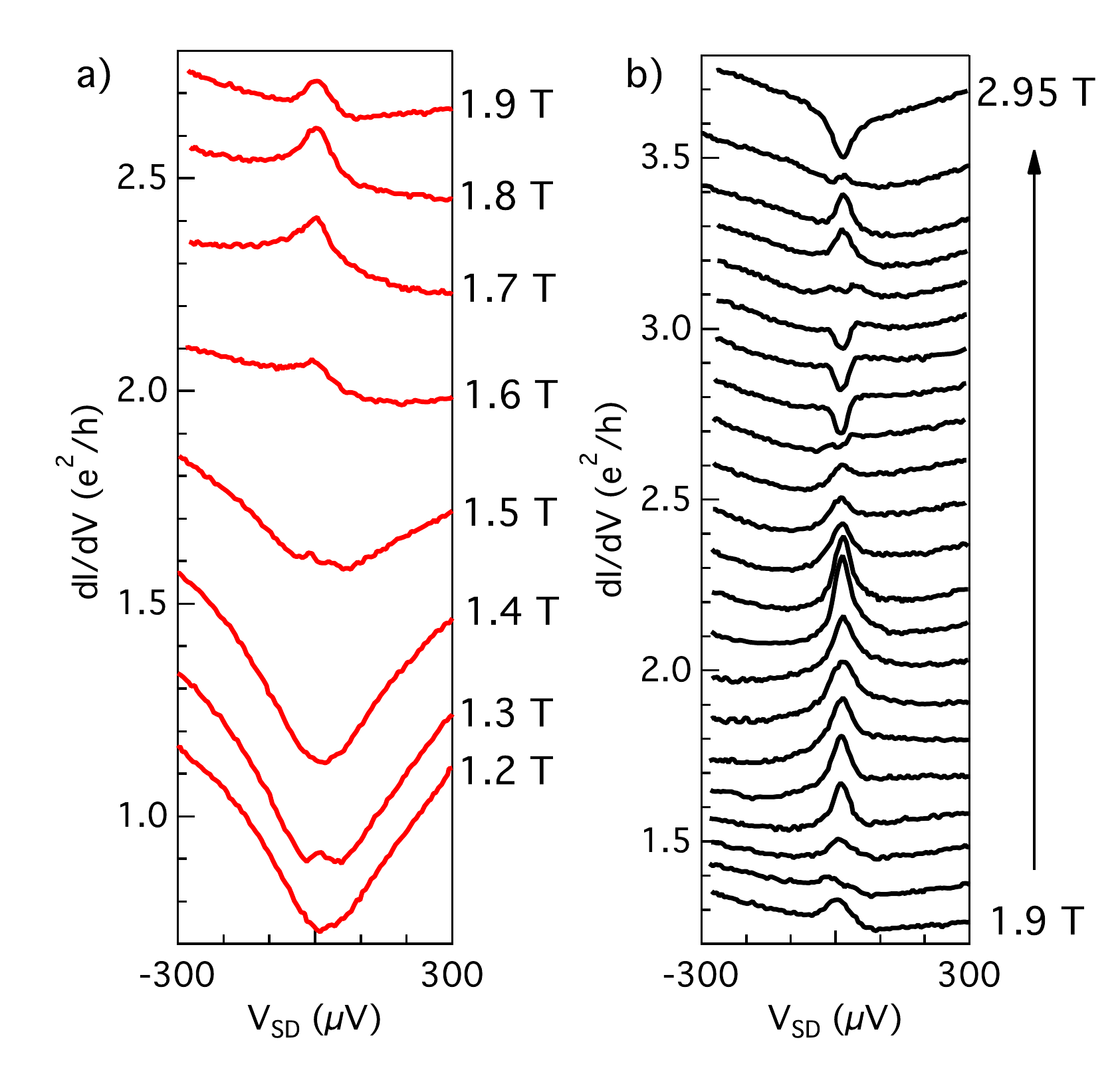}{(a) and (b) Conductance traces at $V_{BG}=10.1$ V for various B values.  In (b), the traces differ by $\Delta B= 0.05$ T.  Adjacent curves are offset for clarity.}

We now consider transport at high magnetic fields.  In Figure \ref{fig:figure2} we show color plots of $dI/dV$ vs $V_{SD}$ and $B$ at two different densities.  For $V_{BG}=10.9$ V, one can observe the zero bias peak at $B=0$ disappear quickly.  Note that the enhancement of conductance for $V_{SD} < 2\Delta_0$ is still present up to $B=3.2$ T, verifying the persistence of the proximity effect.  However, there is no evidence of a ZBA at this density beyond $B=0.4$ T.

The situation is dramatically different for $V_{BG} = 10.1$ V.  Here, the proximity gap seems to close at $B\approx 0.8$ T.  Exceeding this field, the gap first reopens and then gradually closes again.  Beyond $B=0.8$ T, a number of ZBAs are visible.  We repeated the measurement in higher resolution to see the detailed evolution of the ZBA, shown in Figure \ref{fig:figure3}.  The ZBA persists for a range of magnetic field, generally remaining as a single peak at zero energy for several hundred millitesla before splitting.  The ZBA periodically splits and re-forms, with a characteristic interval of $\Delta B \approx 0.6$ T.  A slight disagreement on the location of the ZBA between Figure \ref{fig:figure2}c and Figure \ref{fig:figure3} is likely due to charge noise.

At this point it is tempting to attribute the ZBA to the presence of MFs.  In this picture, the magnetic field drives portions of the nanowire into the topological regime.  The critical field for this transition is a function of chemical potential, thus explaining why no such ZBAs are present at a higher density and why we observe the gap close at lower field ($B=0.6$ T) at an even lower density ($V_{BG} = 9.35$ V).  The topological segments of the nanowire support zero energy end modes.  For example, the regions of the nanowire in direct contact with the NbN leads could exist in the topological regime while the bare nanowire segment, possessing a weaker induced gap, remains trivial.  The MFs at the boundaries between the topological and nontopological segments could then allow the passage of single charges through the bare nanowire segment at zero energy.  Without the MFs, transport is suppressed near zero bias by the superconducting gap, as demonstrated in Figure \ref{fig:figure2}a.

In this interpretation, the periodic splitting of the ZBA would be naturally explained by the hybridization of the MFs.  The Zeeman energy would tune the overlap of the MF wave functions in a periodic fashion, leading to a oscillatory splitting of the ZBA.  For example, Ref.~\cite{PhysRevB.86.220506} calculates the period of this splitting to be $\Delta E_Z = 0.2$ meV for a 1 $\mu$m long topological wire segment.  Determining the Zeeman energy for real devices is difficult due to complications such as confinement \cite{PhysRevB.58.16353} and spin-orbit coupling, both of which can be tuned by external fields \cite{nl802418w, nl901333a, PhysRevLett.107.176811}.  Assuming a value of $g=20$ gives a measured period of $\Delta E_Z = 0.35$ meV, in rough agreement with Ref.~\cite{PhysRevB.86.220506}.

To further test the case for MFs, we explore the behavior of the ZBA with respect to changes in chemical potential.  In Figure \ref{fig:figure4}a, we show a stability plot at $B=2.3$ T.  A ZBA is visible for two different ranges of gate bias.  The persistence of the ZBA with chemical potential is suggestive of a stable set of MFs; their periodic appearance and disappearance would then reflect the gate-tuned hybridization of the MFs.  However, a careful inspection of Figure \ref{fig:figure4}a reveals multiple bands that cross at zero energy where the ZBA occurs.  The separation in back gate bias for this pattern is $\Delta V_{BG} = 0.4$ V, comparable to the periodicity observed at $B=0$.  We posit that this pattern comes from the resonant levels in the nanowire. By crossing at zero energy, these broadened resonances provide the necessary degeneracy to create a persistent ZBA through the Kondo effect.  Indeed, the temperature dependence of the ZBA height in this regime resembles a Kondo effect with a Kondo temperature $T_K \approx 970$ mK \cite{PhysRevLett.81.5225, PhysRevB.84.245316}, as shown in Figure \ref{fig:figure5}.  The resonant levels also evolve with magnetic field; thus the periodic crossing of these levels would explain the modulation of the ZBA in Figure \ref{fig:figure2}b.

\dograph{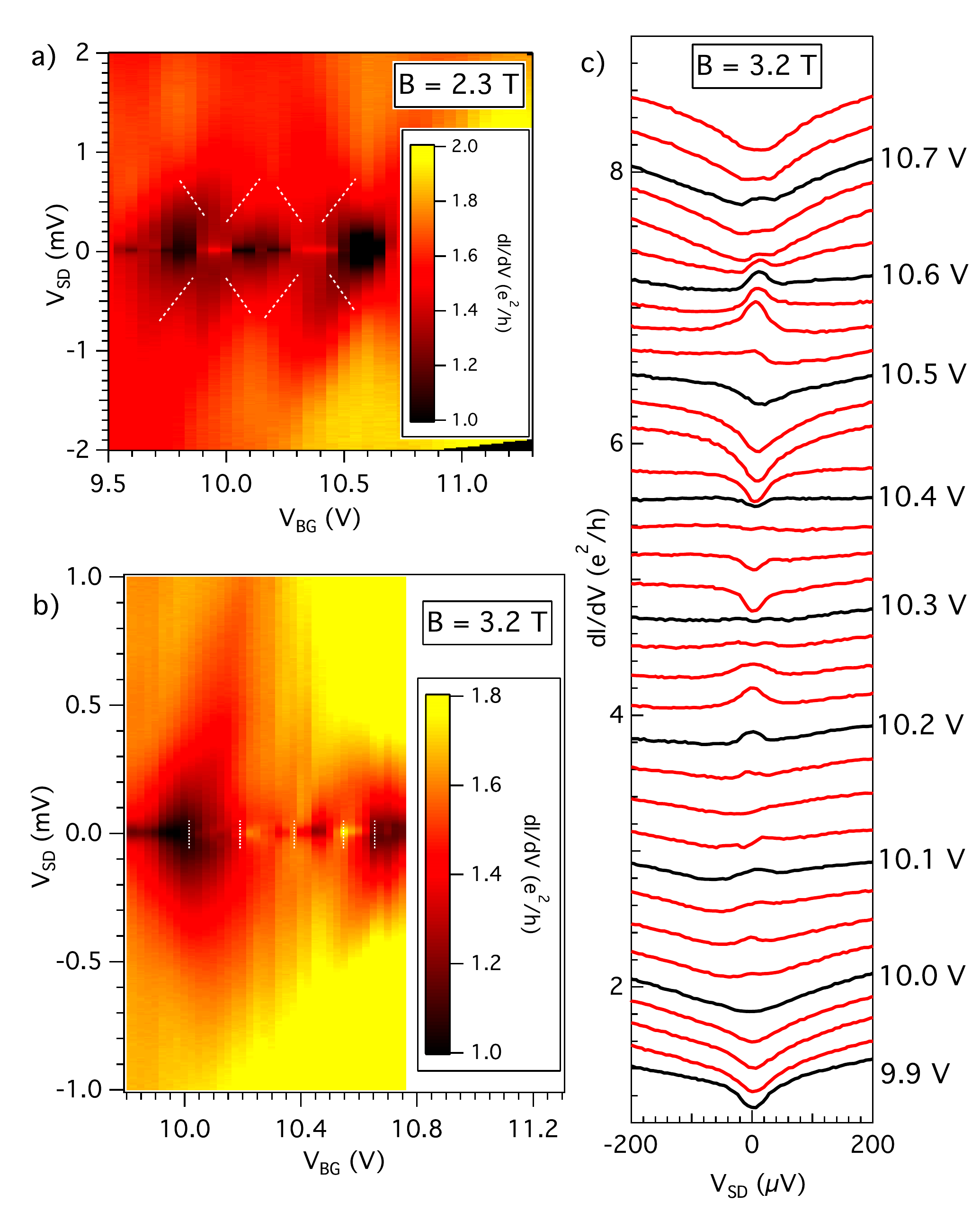}{Stability diagram at (a) $B=2.3$ T and (b) $B=3.2$ T.  In (a), dashed white lines show location of resonant levels.  In (b), dashed white lines delineate individual periods of ZBA oscillations.  (c) Individual traces for $B=3.2$ T in steps of $\Delta V_{BG} = 0.025$, with adjacent curves offset for clarity.}

However, we also see ZBA modulations with frequencies that do not fit a simple picture of regularly crossing resonant levels.  For example, in Figure \ref{fig:figure4}b we show a ZBA that repeatedly splits and re-forms in the range of $V_{BG} = 10$ V to 10.7 V, with a period of $\Delta V_{BG} \approx 0.175$ V.  The observed periodicity of $\Delta \mu = 175$ $\mu$eV is consistent with the predicted MF hybridization period from Ref.~\cite{PhysRevB.86.220506}.  If the ZBAs are caused by the Kondo effect, then this modulation might reflect an RKKY interaction among multiple localized electrons \cite{Craig23042004}.

\dograph{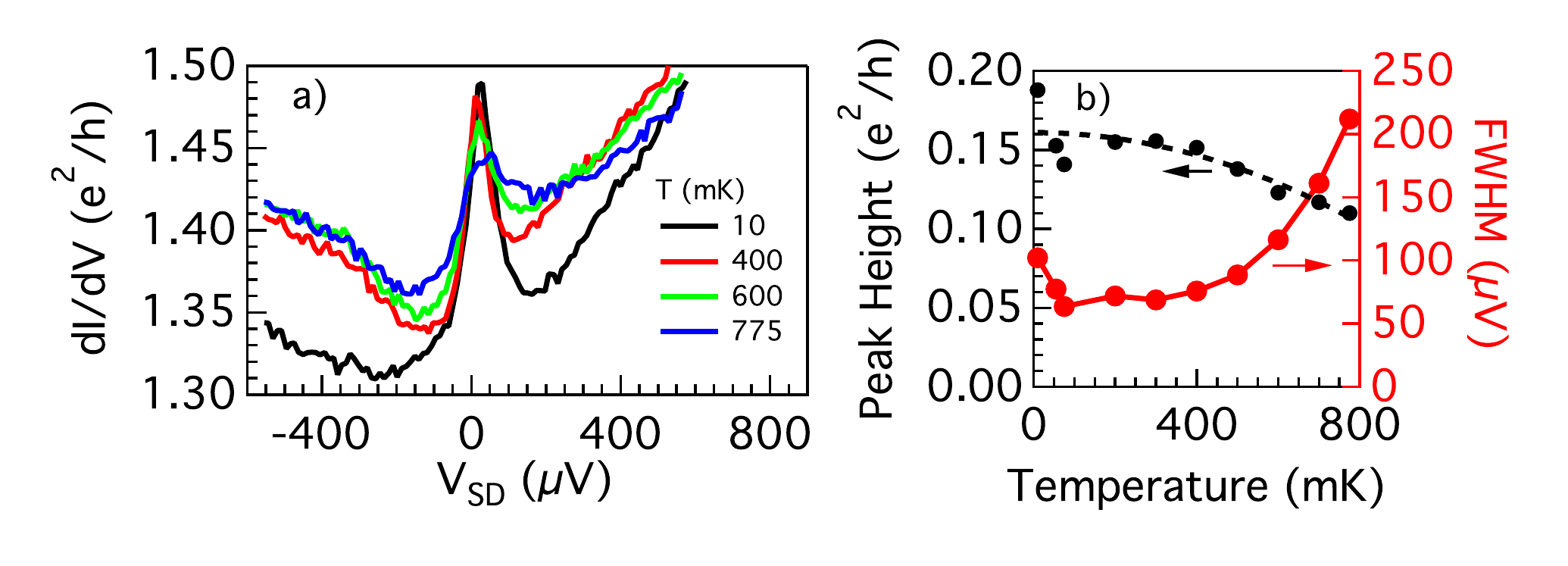}{(a) ZBA temperature dependence at $B=2.75$ T and $V_{BG} = 10.35$ V.  (b) Peak height (black) and FWHM (red) vs T, after subtracting off background.  Black dashed line is parabolic fit based on Ref.~\cite{PhysRevB.84.245316} with $T_K = 970$ mK.}

In conclusion, we observe numerous ZBAs in a nanowire-superconductor device at high magnetic fields.  Their periodic splitting and re-forming are consistent with hybridizing MFs.  However, we also find evidence that the ZBAs result from confined states crossing zero energy and generating a Kondo resonance.

We gratefully acknowledge technical assistance from C.~English, C.~Kurter, C.~Nugroho, V.~Orlyanchik, and M.~Stehno.  We have benefited from conversations with W.~DeGottardi, B.~Dellabetta, S.~Frolov, P.~Ghaemi, M.~Gilbert, T.~Hughes, V.~Manucharyan, D.~Rainis, and S.~Vishveshwara.  A.D.K.F.~ and D.J.V.H.~acknowledge funding by Microsoft Project Q.  P.K.M., K.J., and X.L.~acknowledge funding from NSF DMR under Award Number 1006581.  Device fabrication was carried out in the MRL Central Facilities (partially supported by the DOE under DE-FG02-07ER46453 and DE-FG02-07ER46471).

\bibliography{majoranabib}

\end{document}


\title{Supplementary Materials:\\Anomalous modulation of a zero bias peak in a\\hybrid nanowire-superconductor device}

\author{A.D.K.~Finck$^1$, D.J.~Van Harlingen$^1$, P.K.~Mohseni$^2$, K.~Jung$^2$, X.~Li$^2$}
\affiliation{$^1$Department of Physics and Materials Research Laboratory, University of Illinois at Urbana-Champaign, Urbana, Illinois 61801
\\
$^2$Department of Electrical and Computer Engineering, Micro and Nanotechnology Laboratory, University of Illinois at Urbana-Champaign, Urbana, Illinois 61801}

\date{\today}

\pacs{71.10.Pm, 73.63.Nm, 74.45.+c, 74.78.Na} \keywords{None}

\maketitle

\section{Nanowire growth}

Metal-organic chemical vapor deposition (MOCVD) growth of InAs nanowires (NWs) was performed using a horizontal-flow, low-pressure reactor (AIXTRON 200/4), through the Au-assisted vapor-liquid-solid (VLS) approach. Pre-growth sample preparation of InAs (110) substrates involved a surface degreasing process with standard solvents and native-oxide etching in a dilute hydrochloric acid (HCl) solution, followed by a poly-L-lysine surface coating procedure and substrate decoration with 50 nm Au colloids. Trimethyl-indium (TMI, (CH$_3$)$_3$In) and arsine (AsH$_3$) were employed as sources for the supply of group-III and group-V growth species, respectively, at a constant V/III ratio of 9.8. Nanowire growths were initiated at a reaction temperature of 450 $^{\circ}$C, under a constant 5 L/min flow of hydrogen (H$_2$) carrier-gas, at a system pressure of 200 mBar. In-situ Si doping of InAs NWs was provided through the supply of disilane (Si$_2$H$_6$), in tandem with the flow of TMI.  Nanowire growths were terminated, after a period of 15 minutes, through the simultaneous cessation of TMI and disilane flows. Samples were permitted to cool from the growth temperature under a constant supply of AsH$_3$. Transmission electron microcscopy (TEM) experiments revealed that InAs NWs grown under these conditions exhibit a purely hexagonal (wurtzite) crystal structure, with only several monolayers of zincblende crystal growth at the NW tip. The cubic-phase found at NW tips is attributed to the so-called ``cooling neck" phenomenon \cite{persson2004, plante2008}.


\section{ZBA at low density}

\dograph{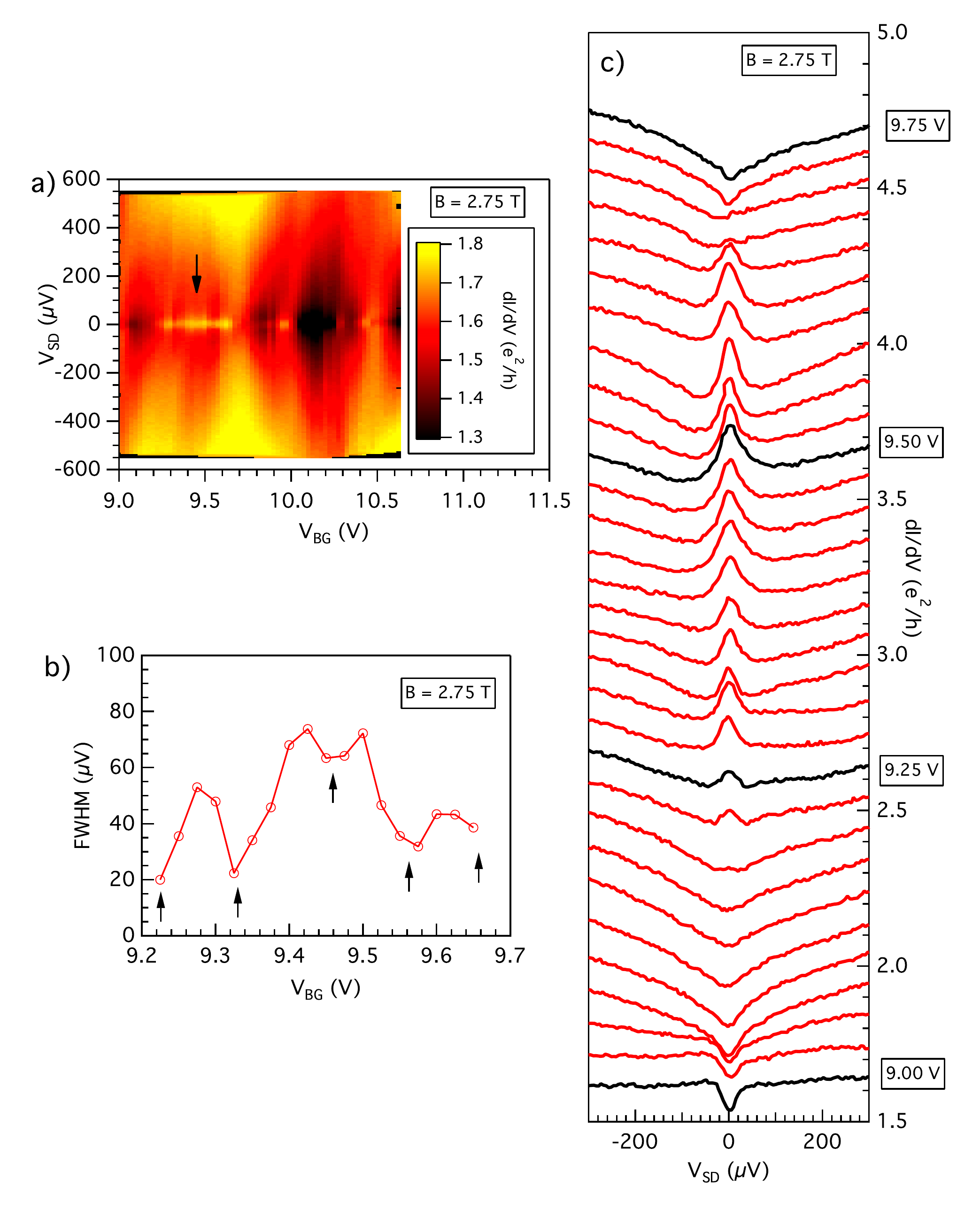}{(a) Stability diagram at $B=2.75$ T.  ZBA is indicated by black arrow.  (b) FWHM of ZBA vs $V_{BG}$.  Arrows point at local minima in oscillations.  (c) Individual conductance traces for region of interest at $B=2.75$ T, taken in steps of $\Delta V_{BG} = 0.025$ V.  Adjacent curves are offset for clarity.}

In Figure \ref{fig:sup_fig2}a, we show the stability diagram for $B=2.75$ T.  A ZBA can be seen in the region of $V_{BG}=9.2$ V to 9.7 V.  We extract the width of the ZBA by fitting to the data a Lorentzian plus an asymmetric v-shaped background.  We plot the FWHM of the fitted Lorentzian vs $V_{BG}$ in Figure \ref{fig:sup_fig2}b.  We see clear oscillations in the width with a period of $\Delta V_{BG} \approx 0.1$ V, which does not match with the spacing of the resonant levels within the exposed nanowire segment.  In Figure \ref{fig:sup_fig2}c, we show the individual conductance traces.

In Figure \ref{fig:sup_fig3}a, we plot the temperature dependence of the zero-bias anomaly (ZBA) for $V_{BG} = 9.45$ V and $B=2.75$ T.  Once again, we determine the height and width of the ZBA by fitting each curve with a Lorentzian and an asymmetric v-shaped background.  Figure \ref{fig:sup_fig3}b shows the height and width of the peak vs temperature.  Note that the width seems to decrease with temperature, in contrast to the expected broadening for a Kondo resonance.

\dograph{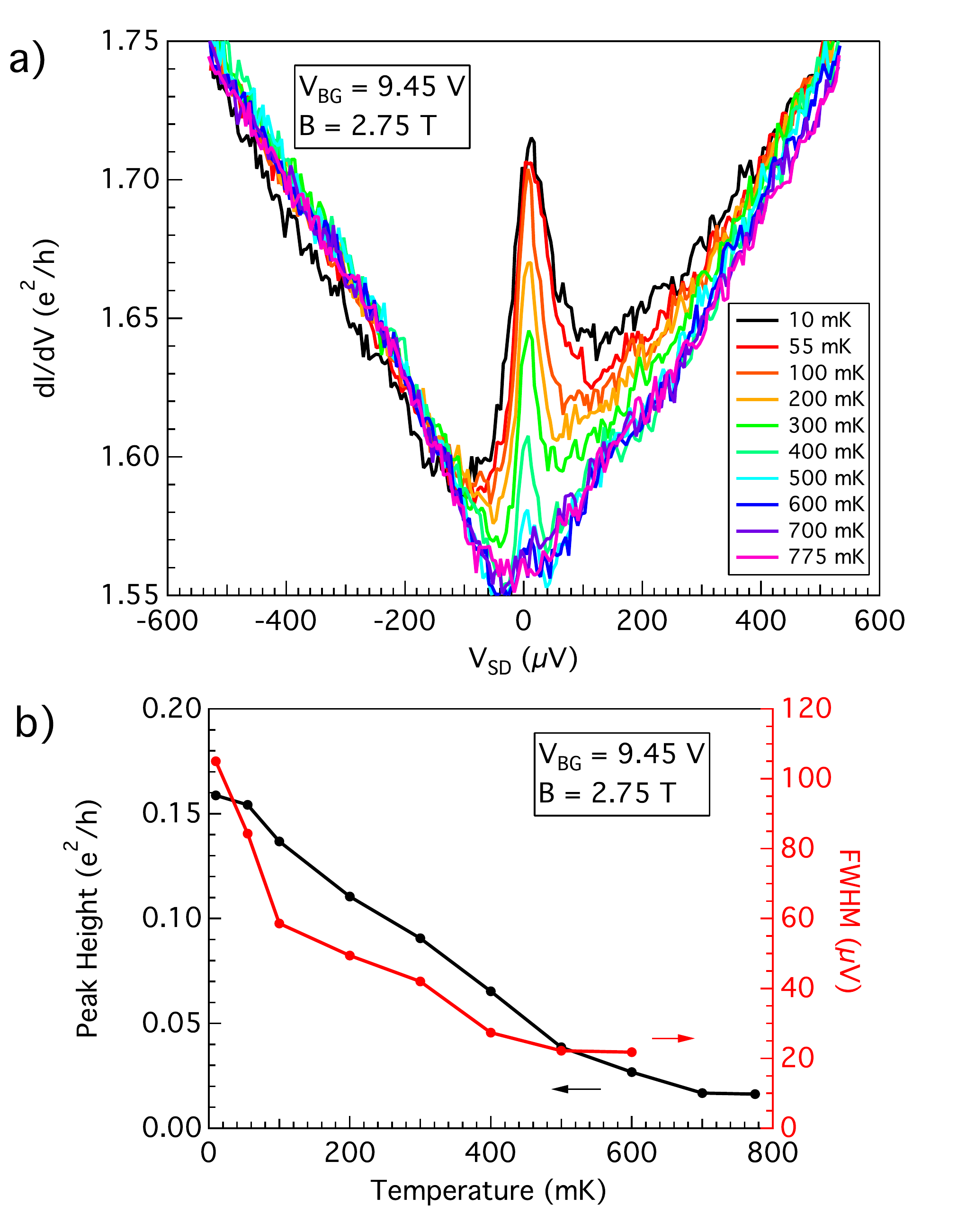}{(a) Temperature dependence of ZBA at $V_{BG} = 9.45$ V and $B=2.75$ T.  (b) Peak height (black trace) and width (red trace) vs temperature.}

